# Detection of low energy single ion impacts in micron scale transistors at room temperature

**A. Batra, C. D. Weis, J. Reijonen, A. Persaud, and T. Schenkel[1]**

Accelerator and Fusion Research Division, Lawrence Berkeley National Laboratory, Berkeley, CA 94720, USA

**S. Cabrini**

The Molecular Foundry, Lawrence Berkeley National Laboratory, Berkeley, CA 94720, USA

**C. C. Lo, J. Bokor**

Department of Electrical Engineering and Computer Science, University of California, Berkeley, CA 94720, USA

We report the detection of single ion impacts through monitoring of changes in the source-drain currents of field effect transistors (FET) at room temperature. Implant apertures are formed in the interlayer dielectrics and gate electrodes of planar, micro-scale FETs by electron beam assisted etching. FET currents increase due to the generation of positively charged defects in gate oxides when ions ($^{121}$Sb$^{12+, 14+}$, Xe$^{6+}$; 50 to 70 keV) impinge into channel regions. Implant damage is repaired by rapid thermal annealing, enabling iterative cycles of device doping and electrical characterization for development of single atom devices and studies of dopant fluctuation effects.

---

[1] Electronic mail: T_Schenkel@lbl.gov

Implantation of dopant ions into FET channels is a standard technique for control of key transistor performance properties, such as the threshold voltage, $V_{th}$. Statistical variations of dopant numbers can lead to large swings in threshold voltages for small devices [1]. Control of FET channel doping by single ion implantation [2-4] and retention of single dopant atom positions through optimized process control [4, 5] enables systematic investigation of single dopant effects [6], and the development of a new class of devices where function is based on the presence of single dopant atoms and on the manipulation of their quantum states [7]. Donor electron and nuclear spins are promising candidates for implementation of quantum bits in silicon [7]. The detection of low energy single ion impacts for device integration has been accomplished via detection of secondary electrons [2, 4, 9], or by collection of electron hole-pairs in optimized diodes [3]. It is also well known that high energy (MeV) single ion impacts can upset device currents [10], and an extension of this approach to low energy ions was recently outlined in Ref. 11. Also, random telegraph noise due to switching occupancies of single Coulomb scattering centers [12] has long been observed in sub-micron transistors, and it can thus be expected that the impact of lower energy (<100 keV) single ions, which is accompanied by the generation of multiple charged defects, can also be sensed in FETs.

In this letter we report on the detection of low energy (50 to 70 keV) antimony and xenon ion impacts in FETs with channel areas of 4 $\mu m^2$ at room temperature. FETs were formed for development of single donor spin readout techniques, and spin dependent neutral donor scattering was recently observed in transport studies with similar devices used here [13]. Single ions change transistor channel mobilities through formation of defects upon impact, enabling precision placement of defined numbers of dopants into transistor channels.

FETs and were fabricated on natural silicon (100) wafers with undoped (n-type, >1 kOhm cm) substrates for formation of accumulation mode (A-FET), and p-type (~1 Ohm cm) substrates for enhancement mode (NMOS-FET) devices, respectively. A conventional Local Oxidation of Silicon (LOCOS) process was used to define channel areas of 2 µm length and width. A 20 nm gate oxide was grown and *in-situ* phosphorus doped poly-silicon was deposited and patterned as the gate electrode with a thickness of 160 nm. A high dose arsenic implant ($5\times10^{15}$/cm$^2$, 40 keV) was then performed to form degenerately n-type doped source/drain regions. Low-temperature chemical-vapor deposited silicon dioxide (LTO) was used as an interlayer dielectric layer with a thickness of 300 nm, contact regions were etched and tungsten was sputter deposited to complete device metallization. An $N_2/H_2$-forming gas anneal at 400º C for 20 minutes was performed to passivated defects at the Si/SiO$_2$ interface and to improve the metal-semiconductor contact quality. Following electrical testing, devices where processed in a dual beam Focused Ion Beam (FIB) system. Here, apertures with areas of 0.1 to 1 µm$^2$ where opened in the passivation layer and poly-silicon gate to allow implantation of low energy dopant ions into transistor channels. First, a 30 keV Ga$^+$ ion beam was used to remove parts of the LTO layer. The remaining LTO was removed by electron beam assisted etching with 5 keV electrons and XeF$_2$ [14]. Following removal of the LTO and parts of the poly-silicon layer, the electron beam was turned off and etching by XeF$_2$ gas alone lead to the formation of apertures in the poly-silicon gate. Since XeF$_2$ does etch silicon but not SiO$_2$ [15] the gate oxide acted as an effective etch stop for this process.

Following FIB processing, devices underwent another forming gas anneal at 400º C for 30 min. Electrical testing validated device integrity, and FETs were then mounted in our setup for ion implantation with scanning probe alignment [16]. Figure 1 shows an *in situ* scanning force image [16] of an A-FET with source, drain and gate electrodes as mounted in the implant chamber. Here, ions were delivered to the target devices from an Electron Cyclotron Resonance source (ECR) or an Electron beam Ion Trap (EBIT). The ECR [17] formed beams of multiply charged xenon ions (e. g. 6+, 50 keV), which are used for device setup and test exposures. The EBIT source [18] delivered beams of antimony ions ($Sb^{7+ \text{ to } 26+}$, $E_{kin}$=35 to 130 keV). Antimony donors are attractive electron and nuclear spin qubit candidates in silicon, since straggling in the implantation process is much smaller then for phosphorus, and because antimony is a vacancy diffuser, which does not segregate to the Si-$SiO_2$ interface upon annealing [5]. Coherence times of implanted $^{121}$Sb donors [5] and gate modulation of their hyperfine coupling have been quantified [19], and spin dependent transport was recently established as a promising mechanisms for projective measurements of single nuclear spin states with implanted $^{121}$Sb donors [13] in A-FETs formed in the same process as the ones used here. Qubit integration calls for a donor depth of about 10 to 20 nm below a gate dielectric [7]. From simulations of implant profiles and extrapolation from previous Sb implants [5], we estimate that this can be achieved with an implant energy of about 60 keV, where the projected range in $SiO_2$ (10 nm)/Si is about 20 nm.

FETs were exposed to ion beams at a pressure of $10^{-7}$ Torr at room temperature. Ions with a desired mass/charge ratio where selected in a 90º analyzing magnet. Ion beam currents of ~ 1 pA/mm$^2$ where tuned so that at average less then one ion/s would hit the ~1 μm$^2$ implant apertures. Ion beams where then pulsed with variable on/off times for monitoring of channel current changes induced by (single) ion impacts.

Figure 2 a) shows the channel current, $I_{SD}$, as a function of time during a 200 s long, pulsed exposure. The device was an A-FET operated at $V_{gate}$= 1.1 V and $V_{sd}$=0.1 V. The source was grounded and the drain current was recorded with a digital oscilloscope as a function of time after amplification in an inverting current amplifier (SR 570). During most exposure intervals of 2 s, a clear step in the channel current is detected. The incident ions where $^{121}Sb^{14+}$ with a kinetic energy of 70 keV. $I_{SD}$ was found to increase by ion impacts for both device types, A-FET and NMOS-FET, and we attribute this to the formation of positively charged defects in the gate oxide [20]. Upon reduction of the ion beam current, intervals with ion hits and no-ion hits, or misses, appear in the time traces of the channel currents. Figure 2 b) shows an example of a time series with hits and misses for $^{121}Sb^{12+}$ ions at 60 keV from an NMOS-FET, together with the drain current noise when the ion beam was blocked by the cantilever of the scanning probe. Upon further reduction of the beam current to ~0.1 ions/s, pulses contain mostly no ions, and current steps from single ion hits are recorded (Figure 2 c)). The probability for multiple ion hits in one pulse under these conditions of reduced beam current was less then 3%.

During exposures with ions of different impact energies and charge states we found that the sensitivity to ion impacts, i. e. the magnitude of current steps, was gradually reduced with increasing implant dose. Further, variations in step heights at the given noise level did not allow us to confidently discriminate multiple hits from single ion hits based on the step heights. Due to the degrading sensitivity, it was also difficult to investigate charge state effects on the single ion induced current step height. It can be expected that the localized deposition of potential energy of multiply and highly charged ions [21] contributes significantly to the formation of defects in the gate oxide and at the Si-SiO$_2$ interface, and future work aims at quantifying this effect.

Following a series of exposures with an accumulated dose of ~$10^{11}$ cm$^{-2}$, devices were annealed for damage repair and dopant activation. Rapid thermal annealing (RTA) was performed in an AGA Heatpulse at 900º C for 20 s in Argon, followed by another 30 min. N$_2$/H$_2$-forming gas anneal at 400º C. In figure 3, we show a series of I-V curves of a pristine A-FET (Fig. 3 a) and NMOS-FET (Fig. 3 b), after FIB processing and forming gas anneal, and then after monitored implantation with noble gas and Sb ions and the consecutive anneals, demonstrating that devices were functional transistors after the full process sequence. The threshold voltages, $V_{th}$, were found to be reduced during the implant process, from a pre-FIB value of -0.2 V, to -1.3 V (for the A-FET) after the Sb exposures and anneals. This can be attributed to the generation of positively charged defects in the gate oxide [20], where the accumulated positive oxide charges produce an effective gate voltage increase. Oxide defects are repaired or passivated in the annealing steps after monitored implantation. The effective dopant dose of ~$10^{11}$ cm$^{-2}$ was too low to observe shifts in $V_{th}$ that could be pinned to activated donors, and $V_{th}$ was found to increase following implantation and annealing for both device types. The processes outlined here allow the preparation of devices where controlled numbers of dopant atoms are introduced into device channels. In smaller FETs, current changes from single ion impacts can be expected to be proportionally larger, as a larger fraction of the channel current is affected by single ion impact induced defects.

In conclusion, we report detection of single ion impacts from low energy dopant ions in micron scale transistors at room temperature. Implant apertures were formed in interlayer dielectrics and gate electrodes of planar FETs by electron beam assisted etching with XeF$_2$. Together with tungsten based device metallization this process enables repeated cycles of ion implantation and rapid thermal annealing, as well as "retrofitting" of functional transistors with specific channel implants [22]. Sensitivity to single ion impacts is demonstrated without device cooling and for relatively large devices and allows extension to doping and transport studies of sub-micron devices in future work.

**Acknowledgments**

Work at Lawrence Berkeley National Laboratory (LBNL) was supported by the National Security Agency under contract No. MOD 713106A, by NSF under Grant No. 0404208, and by the U. S. Department of Energy under Contract No. DE-AC02-05CH11231, including portions of this work that were performed at the Molecular Foundry, LBNL.

**Figure captions**

Figure 1. *In situ* scanning force microscope image of an A-FET with etched hole.

Figure 2: Source-drain currents, $I_{SD}$, as a function of pulsed exposure time. The ion beam is on during pulses indicated by the vertical lines. A) $^{121}Sb^{14+}$ ions (70 keV), B) $^{121}Sb^{12+}$ ions (60 keV), after 10 s, the beam current was reduced to ~1 ion/pulse of 0.5 s. The lower curve shows the channel current noise when the beam was blocked. C) $Xe^{6+}$ (50 keV), with ion beam current reduced to 0.1 ions/s.

Figure 3: I-V curves of an A-FET (a) and an NMOS-FET (b), for pristine devices, after FIB processing and forming gas anneal, and after exposure to noble gas and Sb-ions followed by RTA and another forming gas anneal. The source – drain bias was 1 V.

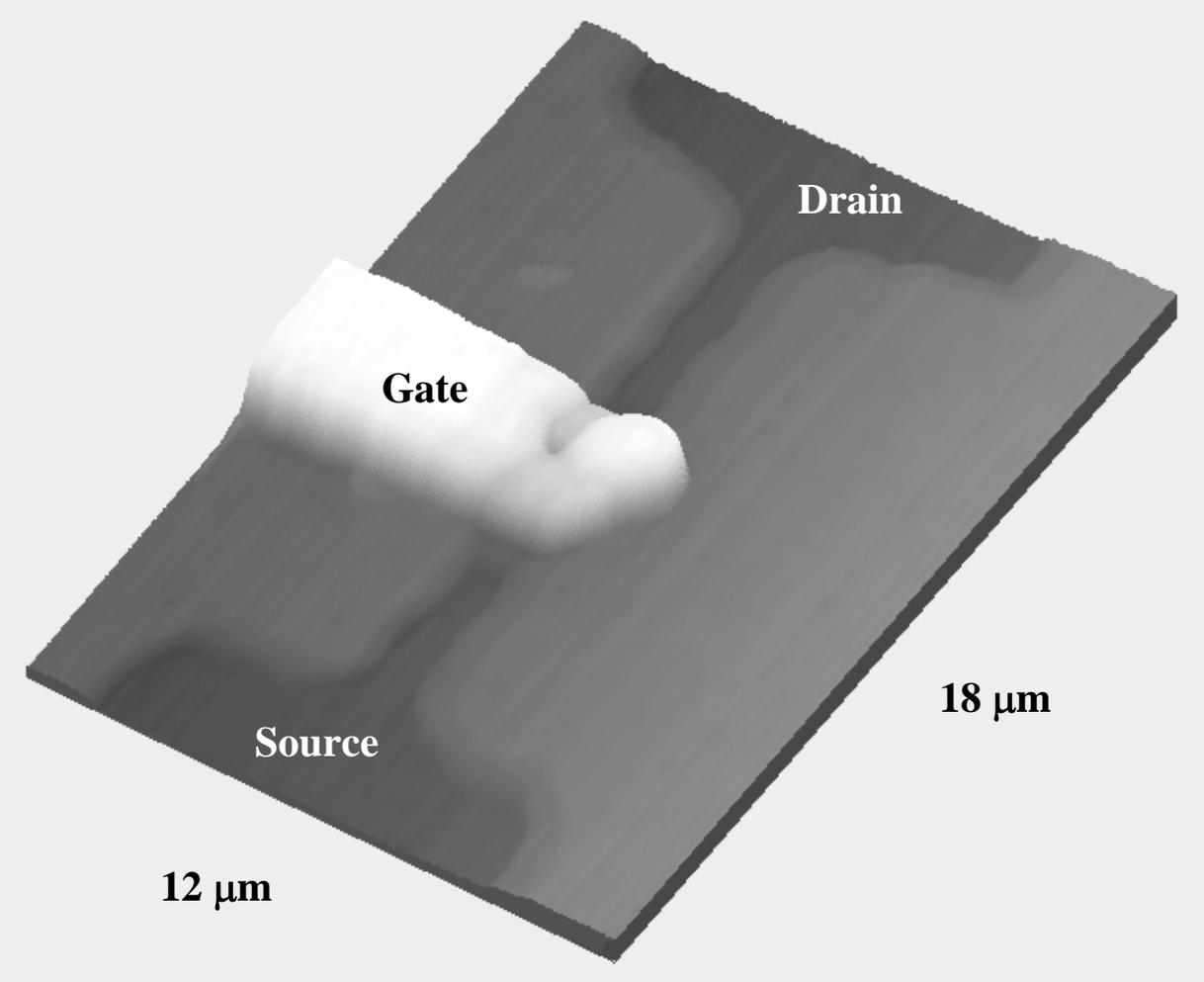

Figure 1

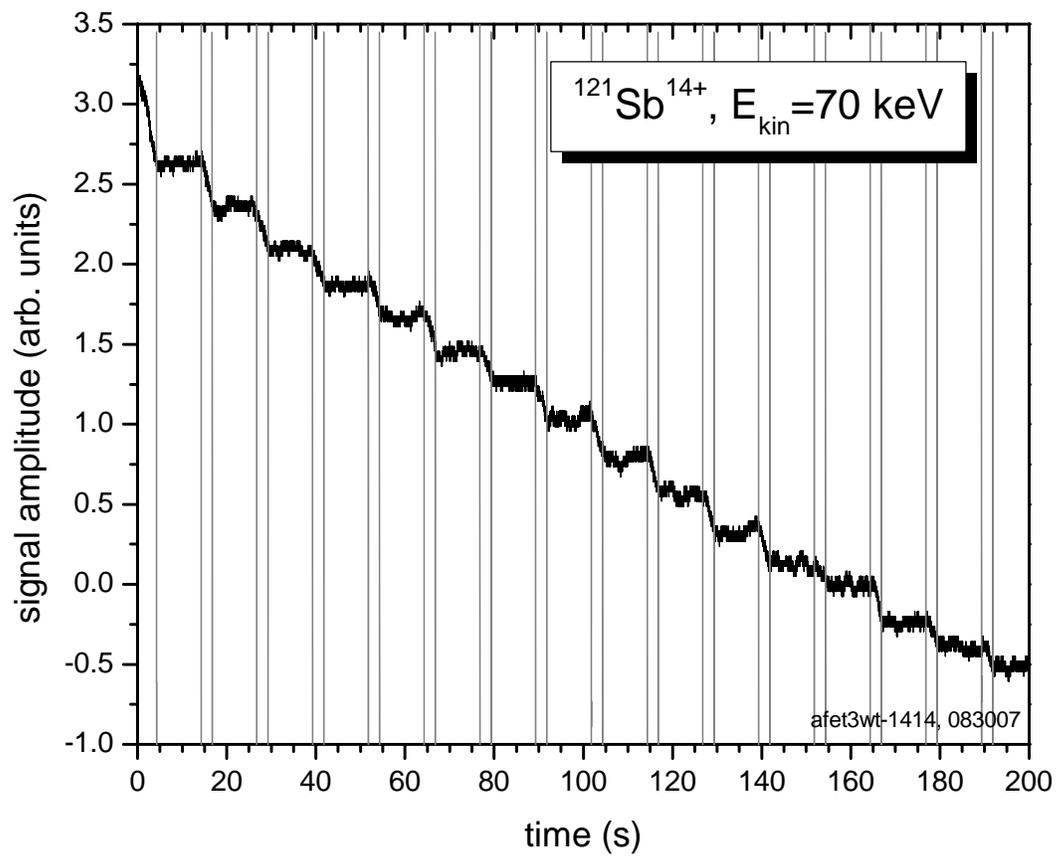

Figure 2 a)

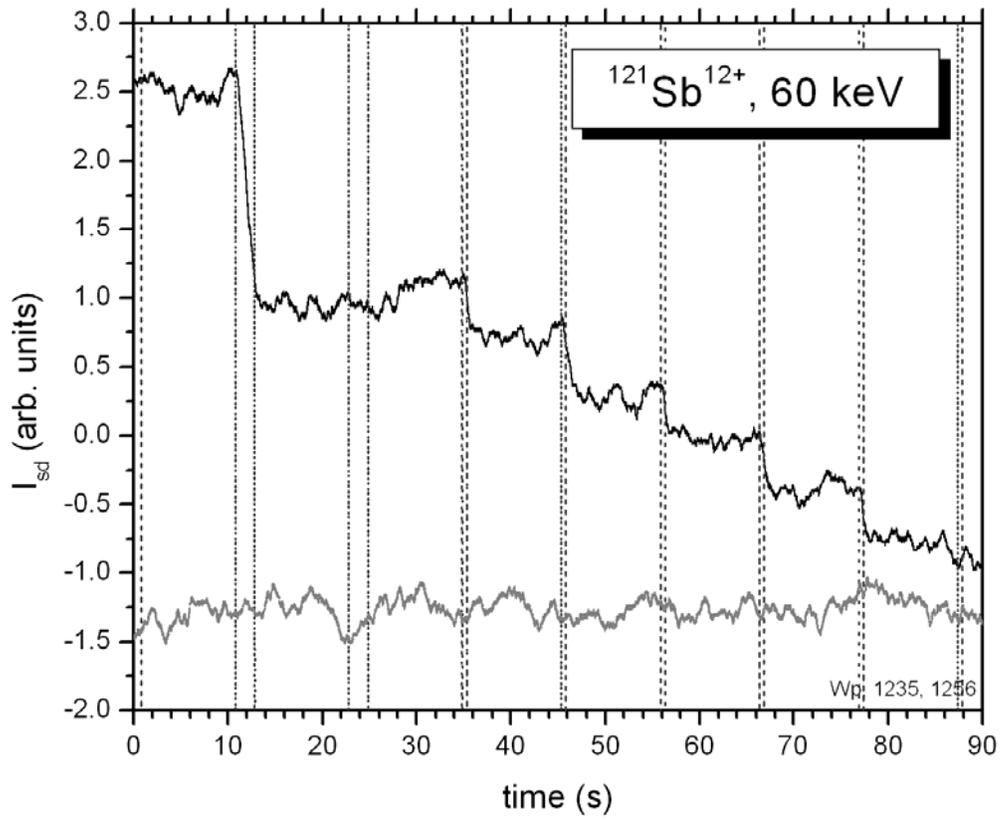

Figure 2 b)

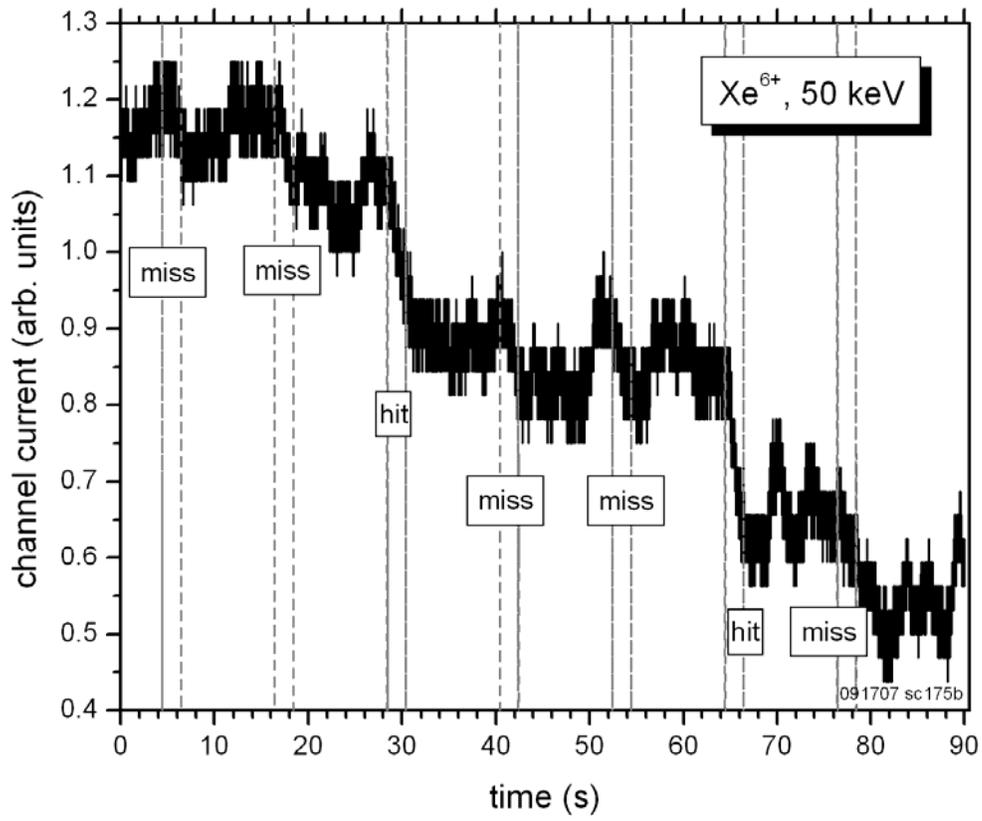

Figure 2 c)

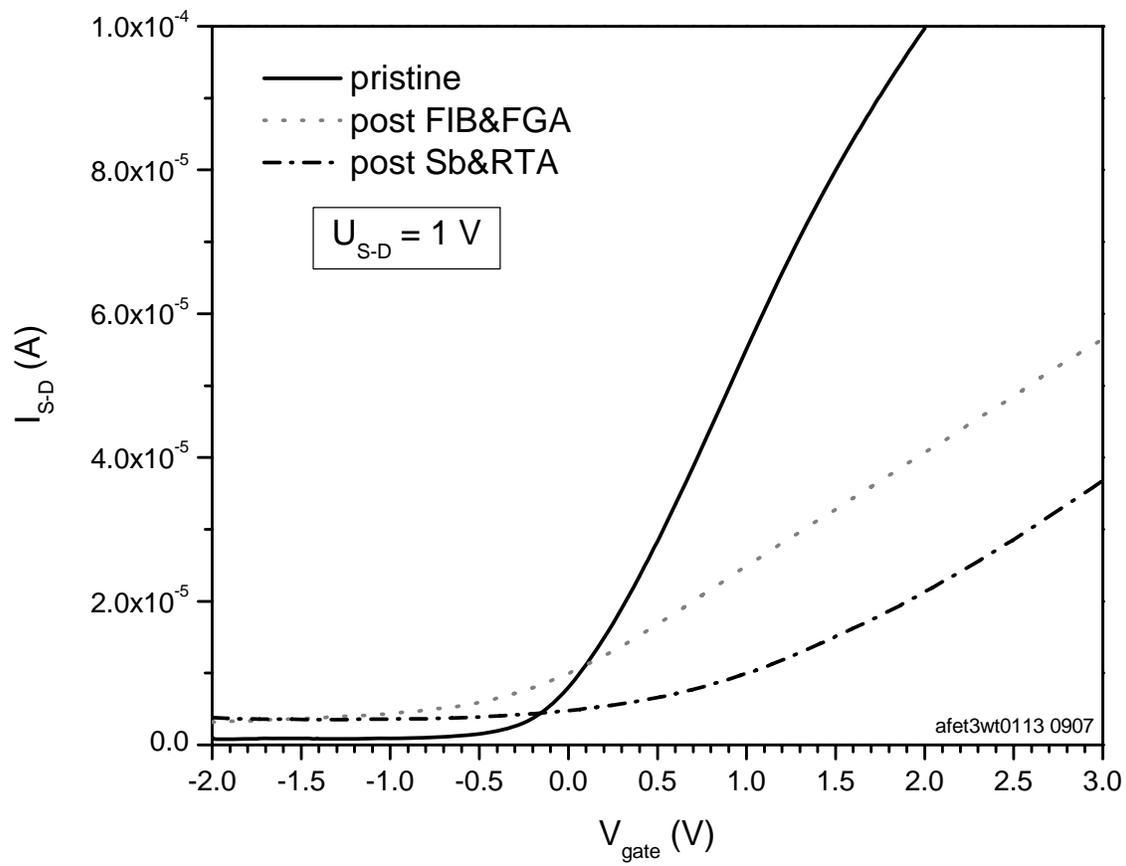

Figure 3 a)

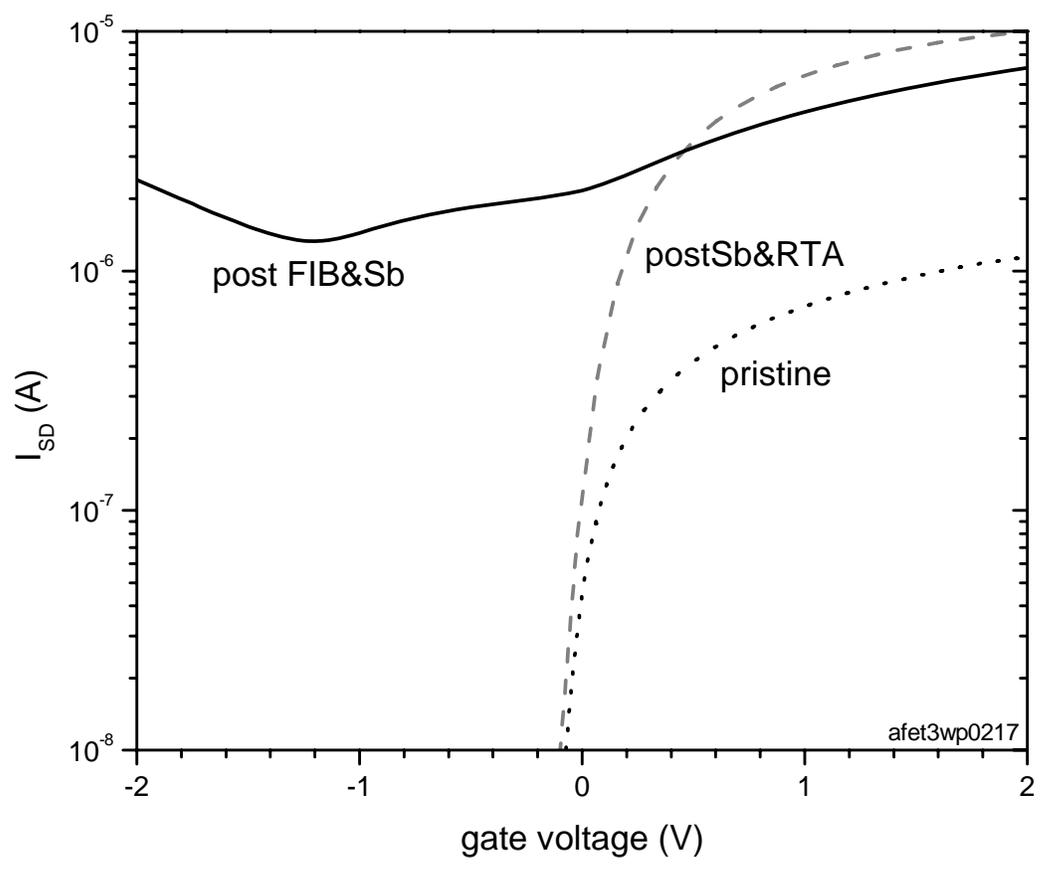

Figure 3 b)